# A novel architecture for DAQ in multi-channel, large volume, long drift Liquid Argon TPC

S. Centro, G. Meng, F. Pietropaolo, S. Ventura

*Università di Padova e INFN Sezione di Padova, via Marzolo 8, I-35131 Padova, Italy.*

Recently a large interest has been shown for multi-ton Liquid Argon Time-Projection-Chambers (LAr-TPC). The physics issues are very challenging, but the technical problem of long drifts and adequately long life-time of free electrons are not solved at all so far. Also one should take into account the extremely large number of channels required for such large volumes. In this paper we propose an architecture for DAQ that is based on recent developments in consumer electronics that made available, at a quite interesting price, components aimed to high-resolution delta-sigma conversion. This type of ADC is not at all popular in HEP experiments where normally signals related to "events", well defined in time (triggered), should be converted and recorded. In the LAr-TPC however we have to deal rather with waveforms that should be converted and recorded continuously, that is the paramount case of delta-sigma ADC application.

## 1. DELTA-SIGMA APPROACH

The delta-sigma conversion technique has been in existence since many years, but only recently the enormous progress in the integration of complex functions on silicon has made it extremely appealing. The main applications are found in the consumer and industrial domain: audio applications, process control, transducers, motor control, industrial instrumentation etc.

ADC resolution beyond 16 bit, up to the limit of 24 bit, is only conceivable with the delta-sigma approach.

The main feature of this technique is to move into digital domain all the conversion process at a very early stage and to exploit the use of numerical digital filtering techniques. The final quality of the converted data is highly dependent on the sampling frequency and numerical filtering.

The basic principle of a delta-sigma modulator is given in Figure 1 [1].

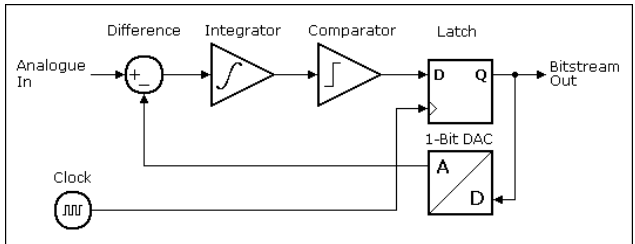

Figure 1: First order delta-sigma modulator.

The analogue part, forgetting the front-end amplifier if required, is made of a very simple comparator, i.e. a single bit ADC that has only two outputs in the signal range with a perfect linearity. This perfect linearity is basically the main reason that allows to reaching the high resolution of delta-sigma converters, unattended by any other technique.

The latch is clocked at the modulation clock frequency, and for a given constant input signal, a periodic sequence of 0 and 1 is obtained. The duty cycle of the output digital value will change if the input signal changes.

As an example, Figure 2 [1] shows the behavior of the delta-sigma modulator components to a sinusoidal input signal.

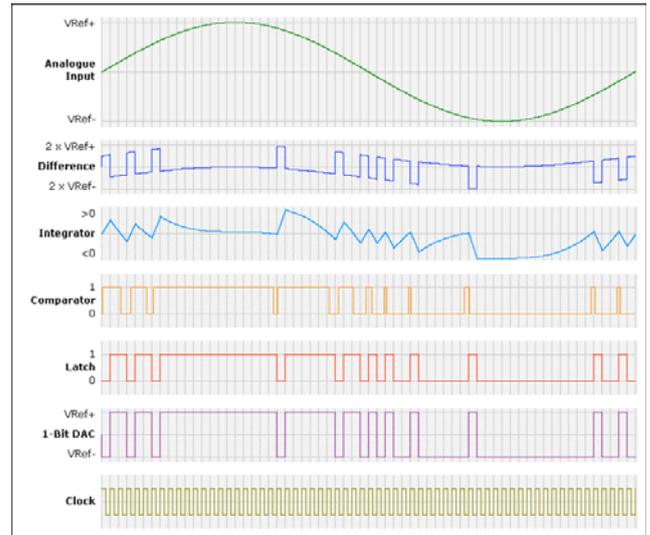

Figure 2: Response of the first order delta-sigma modulator to a sinusoidal input signal.

To better understand the behavior of the modulator in the frequency domain we can use the equivalent circuit of Figure 3 where we use the L-transformation, being $X(s)$ the input signal, $H(s)=1/s$ the transfer function of the integrator, and $N(s)$ the noise source due to the quantizer. Assuming for a moment a noiseless quantizer we obtain

$$[X(s) - Y(s)] * \frac{1}{s} = Y(s) \qquad (1)$$

that brings to the signal transfer function

$$H_{signal} = \frac{Y(s)}{X(s)} = \frac{1}{s+1} \qquad (2)$$

Assuming now that only the quantizer noise and no signal are present at the input we obtain

$$[0 - Y(s)] * \frac{1}{s} + N(s) = Y(s) \qquad (3)$$

that brings to the quantizer noise transfer function

$$H_{signal} = \frac{Y(s)}{N(s)} = \frac{s}{s+1} \qquad (4)$$

Reading equations (2) and (4), we can say that the modulator acts for the signals as a low-pass filter, while for the quantizer noise acts as a high-pass filter.

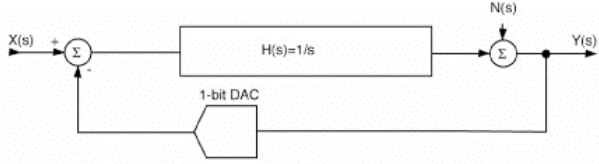

Figure 3: Equivalent circuit of modulator.

The signal after modulation needs to be processed by a numerical digital filter that provides a cut-off at the band of interest and removes as much as possible the (high frequency) quantizer noise.

The quantizer noise is also dependent from the modulation frequency $f_s$ as we will show.

If $q$ is the LSB value (in our case $V_{full-scale}/2$) the relation between the the *rms* quantization noise, $e_{RMS}$, in the conversion interval, and the quantization noise power $e$, are given in the following equation

$$e_{RMS}^2 = \frac{1}{q}\int_{-\frac{q}{2}}^{\frac{q}{2}} e^2 de = \frac{q^2}{12} \qquad (5)$$

Being the maximum frequency $f_0$ of the signal not more than half of the sampling frequency $f_s$ (Nyquist theorem) the sampled signal will have its noise power in the range 0 to $f_s/2$.

The spectral density, $V/Hz^{-2}$, in the maximum allowed range of the signal would be

$$E(f) = e_{RMS}\sqrt{\frac{2}{f_s}} = q\sqrt{\frac{1}{6f_s}} \qquad (6)$$

that shows how it depends on the sampling frequency. More precisely, as the $f_0$ in general should be much lower than $f_s/2$, equation (6) can be converted to in-band (0-$f_0$) quantization *rms* noise ($V^2$) squaring and integrating in the 0-$f_0$ band of interest

$$e_{RMS_{in-band}}^2 = \int_0^{f_0} \frac{q^2}{6f_s}df = \frac{q^2}{12}\left(\frac{2f_0}{f_s}\right) = \frac{q^2}{12}\frac{1}{OSR} = \frac{e_{RMS}^2}{OSR} \qquad (7)$$

where the quantity *OSR* (over sampling ratio) is defined as the ratio between the sampling frequency, $f_s$, and the Nyquist frequency $2f_0$.

Equation (7) summarizes the known result that over-sampling reduces the in-band *rms* noise due to quantization, by the square root of the *OSR*. Doubling the *OSR* decreases the noise by $3dB$.

The output from modulator is passed through numerical filters that perform multistage decimation. The decimation filter(s) reduces the word rate from 1 bit to longer length word: the word length increases to preserve the resolution as the word rate decreases.

The most modern available components, as the low cost ADS1204 from Texas Instruments, use $2^{nd}$ order modulators whose scheme is given in Figure 4 [1]. For this kind of modulator it has been shown [2,3] that equation (7) becomes

$$e_{RMS,in-band}^2 = k\frac{e_{RMS}^2}{OSR^5} \qquad (8)$$

being $k$ about 20, equation (8) shows that, doubling the *OSR*, the in-band *rms* noise decreases by $15dB$. In order to get 10-bit signal resolution, corresponding to a noise attenuation of 60 *dB* (6 *dB* per bit of resolution, $20log_{10}2$), we need an over sampling larger than 16.

It follows that with a $2^{nd}$ order modulator working at 16 *MHz* we can digitize a signal at the rate of 1 *MHz* (Nyquist theorem) with 10-bit resolution (one 10-bit word every $1\mu s$).

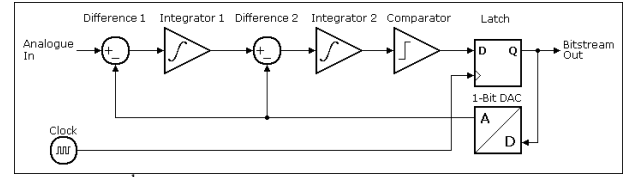

Figure 4: $2^{nd}$ order delta-sigma modulator.

## 2. THE DAQ SCHEME

In the ICARUS T600 LAr-TPC, the AD conversion is performed on sets of 16 channels through a tree of multiplexers and ADCs in an architecture that optimizes trade-off between sampling speed and price. The basic structure is given in Figure 5.

In term of price we can account, only for components, about $30 per 16 channels. The ADCs work at 20Mhz sampling rate, interleaved so the 10bit digital output has a 40Mhz frequency that means that each channel is sampled every 400ns. The power dissipated is significant: 500mW. The required bandwidth taking into account that two sets of 16 channels are merged in a 20 bit word, is 800Mbit/s for 32 channels.

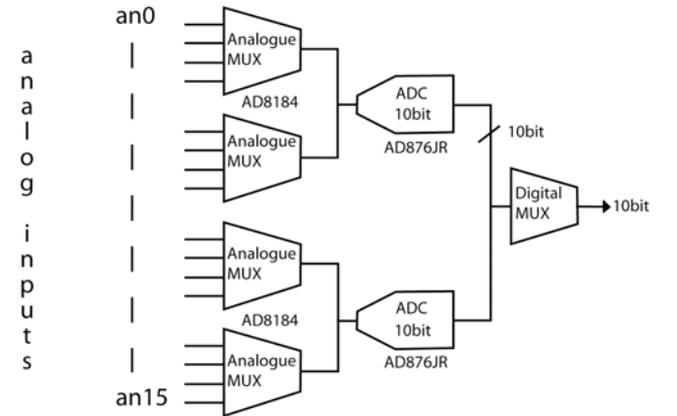

Figure 5: The present DAQ architecture of the ICARUS LAr-TPC.

We have simulated and proposed an alternative structure of conversion based on the delta-sigma serial-converter architecture that has a few interesting characteristics: i) no need of multiplexing, ii) very low number of components, iv) resolution better than 10 bit, if required, iii) low power, and eventually iv) low price.

The basic structure is given in Figure 6 where the serial converter is followed by a simple numerical filter, whose performance will be discussed later, that could be implemented in an FPGA, or in a dedicated DPS. Both solutions could serve much more than 16 channels. Due to the simplicity of the filter, one can even think of leaving the data stream not reduced and make the digital filtering after the digital storage during read out.

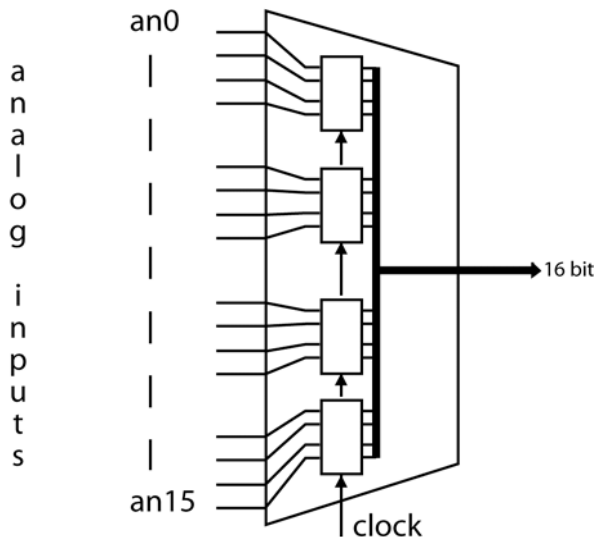

Figure 6: 16 channels DAQ scheme for the LAr-TPC based on the delta-sigma serial converter. In a preliminary prototype the basic component is the commercially available TI-ADS1204.

In a first low cost, low power solution, the circuit is based on the already quoted ADS1204 from Texas Instruments that contains four $2^{nd}$ order delta-sigma modulators working at a maximum frequency of 16*MHz*. This 4 channel chip has a power consumption of only 122*mW* and its cost is as low as 6 $ for modest quantities.

According to what discussed before one can convert signals in the range of 0-500*KHz* with 10-bit resolution. This seems adequate in the case of the LAr-TPC, as shown by the ICARUS results [4,5]. A 10-bit equivalent data will be delivered every 1$\mu s$.

Data could be sent to the computer, not filtered. Through an interface, data from 16 channels will be packed in 16-bit words each containing the *n-th* bit of all the channels, with 256*Mbit/s* data rate.

One important feature of the system is the cost per channel that will be extremely low considering that the modulator function will be less than 1,5$ per channel and AD conversion is fully performed without multiplexing neither analogue nor digital.

A prototype circuit, compatible with the existing ICARUS read-out electronics, is already available (see Figure 7). Both simulations and preliminary tests performed so far give very encouraging results. The circuit will be also tested on a ICARUS LAr-TPC prototype facility.

Note that in case of a delta-sigma modulator sampling at 25*MHz* a 10-bit equivalent data will be delivered every 640 *ns* and the bandwidth for 16 channels would be 400*Mbit/s*. These features are closer to the present ICARUS read-out requirements. With a more performing *OSR* filter (e.g. non linear *FIR* filters), 400*ns* data delivery and 10 bit equivalent data are easily reachable.

The latter solution could be implemented in a circuit based on the existing single channel component, ADS 1604, which features a modulation frequency up to 40*MHz* and integrates a reconstruction FIR filter that allows 16bit analogue-to-digital with sampling rate equivalent to 2.5*Mhz*. The drawbacks of this solution would be the cost (~20$/chip) and the power consumption (0.5*W*/chip).

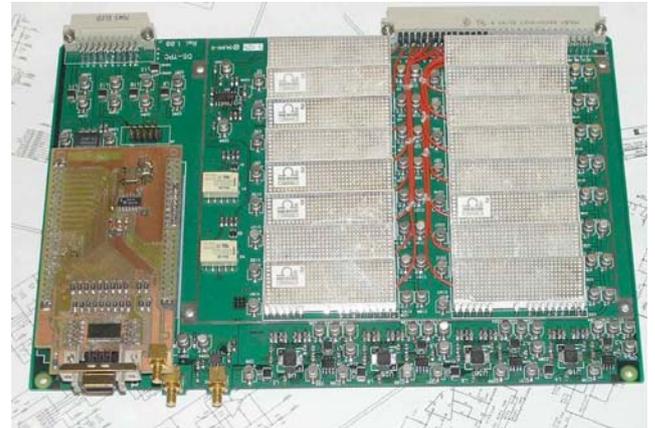

Figure 7: Prototype DAQ board based on the serial converter architecture. Top-right: front-end analogue preamplifiers. Left: digital link. Bottom-right. Delta-sigma converters + gluing logics.

## 3. SINC FILTER

A drastic reduction of the digital noise is obtained with the sinc filter. This choice is justified by the simple structure of this one compared with the more complex structure of the *FIR* filters. The choice of the third order, *L*=3, for the sinc filter is due to the order *K*=2 of the modulator as it has been shown [3] that for *L*=*K*+1 the filter function is close to being optimum for decimating the signal from a modulator of order *K*, working at 16*MHz*

The third order sinc filter, sinc3, has been implemented, in software so far, according to block diagram of Figure 8, with three integrators in cascade (accumulators) followed, after a re-sampling register, by three differentiators. In front of the first accumulator a simple encoder transforms the 1's and the 0's coming from the modulators in two's complement notation.

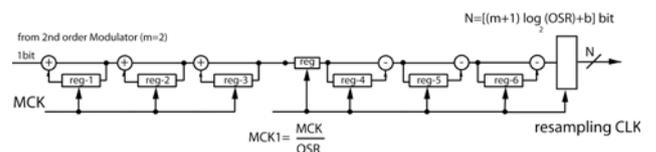

Figure 8: Sinc3 filter.

To prevent overflows, the size $N_{reg}$ of the registers through the whole filter is defined by the following formula [3]

$$N_{reg} \geq (K+1)\log_2(OSR) + b \qquad (9)$$

where $b=1$ for a binary quantizer, $K=2$ is the modulator order, and $OSR=16$. According to this formula 20 bit registers have been adopted. The re-sampling frequency at the output of the sinc filter is $1MHz$ for decimation. According to Nyquist theorem we can represent signals with $500kHz$ bandwidth.

As an example Figure 9 shows the response of the six registers of the sinc3 filter to a step input signal spanning the full dynamic range ($-V_{ref}$ to $+V_{ref}$). As clearly visible, overflows are dealt correctly by the 20 bit registers and the output signal presents a characteristic rise-time, 3 times the OSR window, introduced by the sinc3 filter and due to the triple discreet differentiation (registers R4 to R6), each performed over a width equal to OSR.

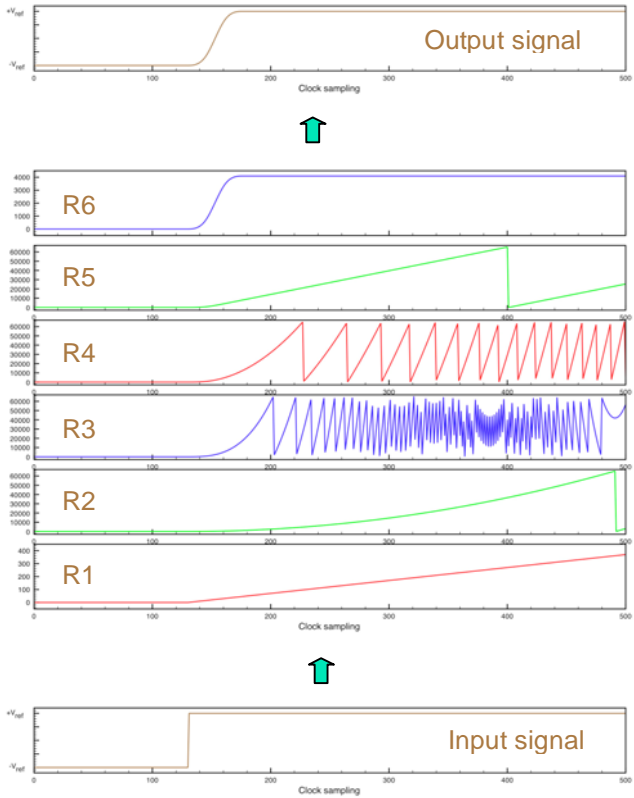

Figure 9: Response of the 6 registers of the sinc3 filter to a step input signal spanning the full allowed dynamic range.

The bit stream produced by modulator treating a real signal from a LAr-TPC, is given in Figure 10, where the first signal in time is the test signal. The delayed signals are those produced by modulator after numerical filtering. The latency depends on the adopted filter scheme.

An interesting feature of this readout system, not available in a classical FADC architecture, is the possibility to reconstruct the data waveform in an almost continuous way (the delayed continuous curve of Figure 10). This feature could help in the data analysis, where one could avoid fitting the data with empirical analytical functions to extract the deposited charge but rather use the numerical integrals of the data themselves.

A trivial modification of the sinc3 filter scheme, allowing the implementation of this useful feature, consists in replicating the differentiation stages N times (with N = OSR) and shifting each one with respect of the preceding by one clock sampling (see Figure 11).

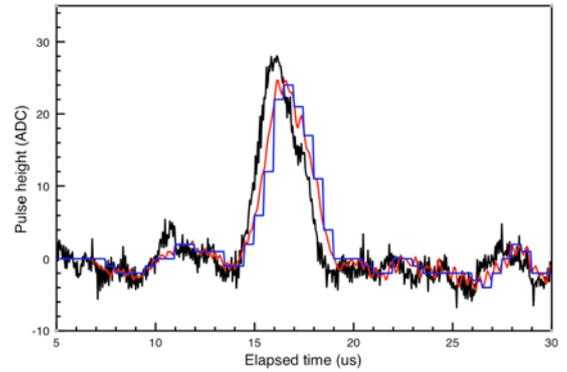

Figure 10: Delta-sigma conversion (16MHz) + Sinc3 reconstruction (OSR=16) of a LAr-TPC signal waveform (from oscilloscope). Black: input signal. Red: continuous signal reconstruction. Blue: decimated output signal.

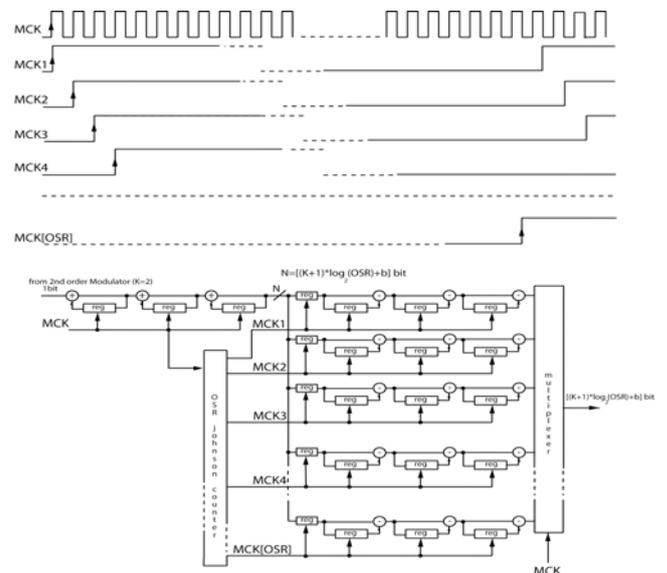

Figure 11: Implementation of the sinc3 filter, allowing an almost continuous reconstruction of the data waveform.

For completeness, we simulated the behavior of the read-out system based on a delta-sigma modulator followed by a sinc3 filter for signal reconstruction, in the case of the fastest ICARUS signals ($1\mu s$ rise-time, $3\mu s$ decay-time), for various combinations of the modulation frequency and over-sampling rate. The result is shown in Figure 12.

The amplitude of the reconstructed signal is smaller than the original one, due to the bandwidth limitation introduced by the delta-sigma modulator. However the area (that carries information about the deposited charge) remains unchanged. It is clear that a system closer to the ICARUS read-out requirements in term of bandwidth and data flow (bottom plot in Figure 12) gives better results.

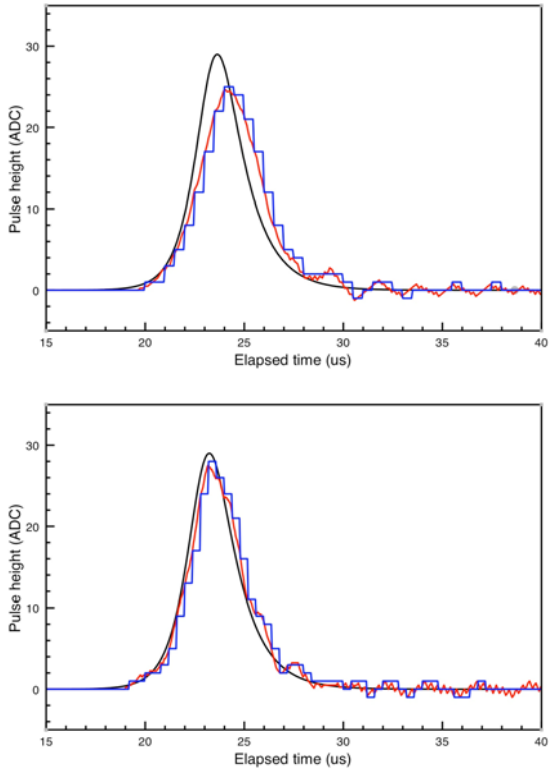

Figure 12: Simulated ICARUS like signals digitized with a delta-sigma modulator (top=16MHz, bottom=25MHz) and reconstructed with a sinc3 filter (OSR=16). Black: input signal. Red: continuous signal reconstruction. Blue: decimated output signal.

These signal distortions are due to the frequency cut introduced by the sinc3 filter working at the delta-sigma modulator sampling rate and with the specified OSR. As an example Figure 13 shows the frequency response in case of 16 and 25 MHz with OSR=16.

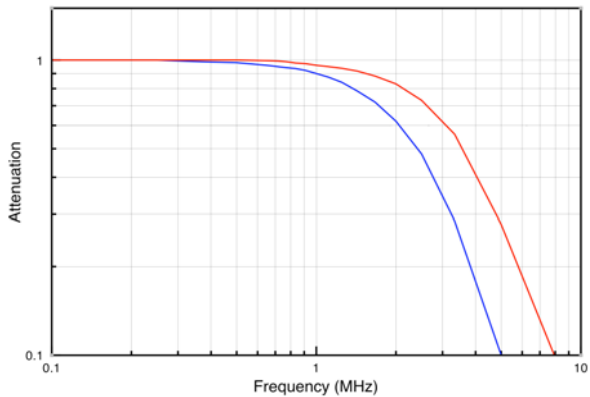

Figure 13: Frequency response of the sinc3 filter in case of 16 MHz (bottom curve) and 25 MHz (top curve) sampling rate and with OSR=16.

Note that the simulation shown in the top picture corresponds to the read-out scheme based on the ADS1204 chip, while that shown in the bottom picture is based on the ADS1602 chip.

Note also the presence of a residual quantization noise, which is, however, well within 1 ADC count even in the low cost solution.

### 4. PRELIMINARY RESULTS

The debugging of the prototype DAQ board described above is underway. For this purpose, it has been included in the ICARUS read-out system and test pulse data, modulated through the delta-sigma converter, have been recorded. A sinc3 reconstruction filter has been applied off-line, according to the scheme of Figure 11. First results are quite encouraging, being both the serial conversion and the sinc3 reconstruction behaving as predicted.

Figure 14 shows a typical test pulse waveform. The high frequency noise visible on the input signal is due to the present board layout (analogue to digital interference).

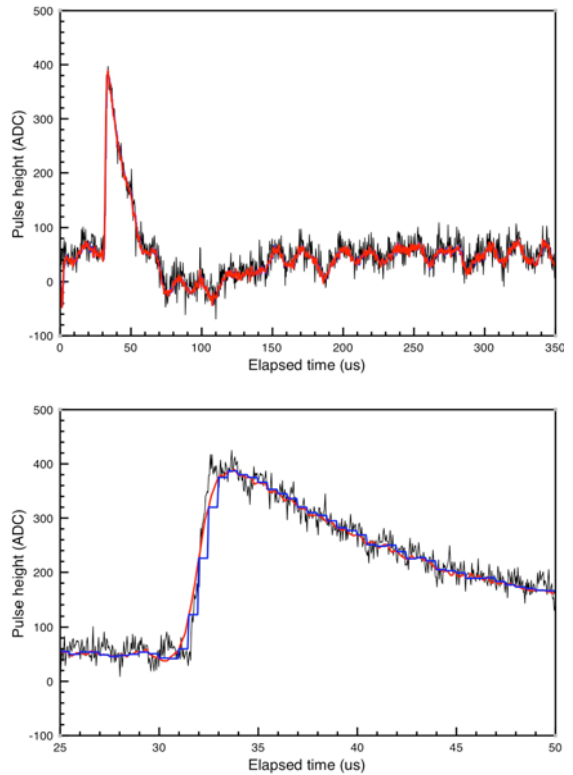

Figure 14: Response to a test-pulse of the prototype DAQ board based on delta-sigma conversion (16MHz) + Sinc3 reconstruction (OSR=16). Black: input signal. Red: continuous signal reconstruction. Blue: decimated output signal. The bottom plot is an expanded view of the top one around the test-pulse region.

A further stage will consist in taking real data with a LAr-TPC chamber and compare the reconstructed track with that collected with the standard ICARUS DAQ architecture, in terms of physical parameters such as energy and space resolution, 3D reconstruction, track separation, etc.

## 5. CONCLUSIONS

A new digitization scheme for the ICARUS read-out has been proposed. It is based on a delta-sigma 1-bit modulator followed by a reconstruction filter and it allows performing digitization of the ICARUS LAr-TPC signals without significant degradation with respect to the existing read-out system. The circuit can be designed around available industrial components. Several solutions are under study: from a low cost, low power one based on the ADS1204 chip, to more performing but more costly and more power consuming based on the ADS1602 chip.

Preliminary results with a prototype board compatible with the present ICARUS read-out system are already quite promising.

One can expect that other components with high frequency modulation, similar to the ADS1602 but without digital filtering, will be sometime available for optimal solution: low power, low cost, high bandwidth and high integration.

## 6. ACKNOWLEDMENT

We warmly thank Uwe Beis that allowed the use of figures part of his web page referred in [1].